\documentclass[a4paper,12pt]{article}
\usepackage{authblk}
\usepackage{graphicx}
\usepackage[margin=1in]{geometry}
\usepackage{graphicx} 
\usepackage[natbibapa]{apacite}
\usepackage{amsmath}
\usepackage{amssymb}
\usepackage{amsmath}
\usepackage{booktabs}
\usepackage{setspace}
\usepackage{hyperref}
\usepackage{subfigure}
\DeclareMathOperator*{\argmax}{argmax}
\providecommand{\keywords}[1]
{
  \small	
  \textbf{\textit{Keywords---}} #1
}

\title{\Large \bf{Massive Retail Location Choice as a Human Flow--Covering Problem}}
\author[1,2]{Hongmou Zhang}
\author[3]{Hezhishi Jiang}
\author[4]{Yihang Li}
\author[4]{Qing Lu}
\author[5]{Yu Liu}
\author[4,*]{Liyan Xu}
\affil[1]{School of Government, Peking University, Beijing, China}
\affil[2]{Institute of Public Governance, Peking University, Beijing, China}
\affil[3]{Academy for Advanced Interdisciplinary Studies, Peking University, Beijing, China}
\affil[4]{College of Architecture and Landscape, Peking University, Beijing, China}
\affil[5]{School of Earth and Space Sciences, Peking University, Beijing, China}
\affil[*]{\emph{Corresponding author (email: \href{mailto:xuliyan@pku.edu.cn}{xuliyan@pku.edu.cn})}}
\date{}

\begin{document}

\maketitle
\begin{abstract}
    In this article we reframe the classic problem of massive location choice for retail chains, introducing an alternative approach. Traditional methodologies of massive location choice models encounter limitations rooted in assumptions such as power-law distance decay and oversimplified travel patterns. In response, we present a spatial operations research model aimed at maximizing customer coverage, using massive individual trajectories as a ``sampling'' of human flows, and thus the model is robust. Formulating the retail location selection problem as a set-covering problem, we propose a greedy solution. Through a case study in Shenzhen utilizing real-world individual trajectory data, our approach demonstrates substantial improvements over prevailing location choices.
\end{abstract}
\keywords{Massive Location Choice, Set-Covering Problem, Spatial Interaction, Shenzhen}
\section{Introduction}
In this article, we offer another look at the classic  business location choice problem, primarily massive location choices for retail chains. In lieu with the widely acknowledged significance of ``location, location, location'' clich\'{e}, the  conventional business location models grounded in classical spatial equilibrium theories, such as those by \cite{von2022isolierte} \cite{christaller1933zentralen}, and   \cite{losch1954economics}. Nonetheless, the conventional approaches encounter notable theoretical and practical challenges, primarily stemming from their predominantly explanatory, rather than normative, nature.

One significant challenge to the conventional model is related to the foundational assumption of a power-law distance decay function, a key assumption in models like the gravity model \citep{reilly1953law}. Recent empirical studies question the validity of this assumption, particularly concerning the arbitrary nature of the commonly presumed power-law decay. Urban environments, far from conforming to the classical ``featureless plain'' assumption \citep{Brueckner1987}, exhibit complex distance decay functions. Efforts to rectify this with multimodal transportation models introduce new theoretical complexities, with implications for both theoretical understanding and computational feasibility.

The oversimplification of travel patterns as a ``residence--store'' two-point linear model is another challenge. Modern cities, marked by mixed-use functionalities, manifest diverse and dynamic travel behaviors. The inherent dynamism of these urban landscapes poses challenges for classical models to capture the nuanced dynamics of store visits effectively.

Polycentrism in cities adds yet another layer of complexity, challenging the applicability of traditional employment location models to the realm of retail. Despite attempts to address decentralization in employment, the unique characteristics of retail location selection, particularly for chain stores, often involve arbitrary pre-specification of city center locations.

Balancing the inherent competition and agglomeration effects further complicates the theoretical landscape. Foundational economic principles supporting agglomeration introduce increased competition, and finding a nuanced equilibrium still remains a theoretical question. Classical theories, though informative, offer limited insights in this agglomeration--competition tradeoff.

Lastly, computational challenges persist. Theoretical issues in business location selection are often difficult to compute and demand high-quality data. On the practical side, simplistic approaches to simulate bid--rent curves through proxies like housing prices and Points of Interest (POIs) lack universality and direct evaluative capacity for location selection.

In response to these challenges, in this paper we advocate for an alternative perspective. Departing from the pursuit of a consistent, analytical distance decay function, we propose viewing the massive retail location selection problem as a spatial operations problem. By optimizing for \emph{maximum customer coverage} and considering constraints related to customer mobility and store characteristics, our approach leverages advancements in spatiotemporal big data technology to address some of the shortcomings of classical models.

This alternative model avoids specific assumptions about distance decay, travel patterns, polycentrism, and competition--agglomeration dynamics, thus allowing for a more pragmatic and adaptable approach to massive retail location selection, capitalizing on observational capabilities afforded by contemporary data technologies.

\section{Literature Review}
In addressing the issue of retail location choice, prior research has developed two primary logics \citep{Marianov2024}: a competition-based one \citep{Hotelling1929,Drezner2024} and a maximum coverage one \citep{Church1974,Garcia2015}.
\subsection{The competition logic}
The location choice research based on competitive logic originated from Hotelling's introduction of consumer location and transportation distance assumptions to address the duopoly problem \citep{Hotelling1929}. Along this logic, researchers take store locations and pricing as decision variables, with  firms' profit maximization as the optimization objective, and consumer store choice behavior and available location options as constraints. 

In the early stages of this theoretical thread, the optimization of the location choice problem was solved using Calculus of Variations \citep{Hotelling1929}, but in later developments, branch-and-bound and heuristic algorithms were widely introduced \citep{Drezner2024}. Building on the Hotelling model, subsequent studies have explored various topics, including the expansion of it into two-dimensional spaces \citep{Drezner1982} or network spaces \citep{Hakimi1983}, introducing more competitors \citep{Brenner2005}, involving consumer behavior in terms of store choice and resource allocation \citep{Fernandez2017,Huff1964} into the model, as well as considering static versus dynamic competition and price elasticity \citep{Eiselt1993}. 

Beyond these variables, competitive strategies and game theory are also considered in this type of research. For example, concepts such as $p$-median and $p$-center have been proposed to address the competition between firms that enter the market sequentially \citep{Hakimi1983}. 

Although these studies have thoroughly considered real-world competition scenarios and generalized the concept of distance to include factors such as product quality and social distance \citep{Hotelling1929}, most of them still adopt a static distance measurement and a static demand point assumption \citep{Drezner1982,Fernandez2017,Hotelling1929}. Specifically, the distance between any consumer and any potential location (and store characteristics) is fixed. This static distance assumption contradicts the reality of consumer mobility. As consumers move, their distances to locations can change, which in turn affects transportation costs to stores and purchase decisions. Relaxing this static assumption remains a research gap.

\subsection{The maximum coverage logic}
The location choice research based on maximum coverage logic approaches the problem through the coverage relation between \emph{potential locations} and \emph{demand points}, thus typically formulating into an integer programming problem  \citep{Church1974}. This type of research can take various forms, such as minimizing the total distance between facilities and demand points or ensuring complete coverage of demand points within a specified distance threshold  \citep{Plane1977}. Over time, maximum coverage location choice research has adopted various constraints, objectives, and model designs to address multifaceted real-world scenarios, such as capacity constraints \citep{Pirkul1989}, queue times \citep{Marianov1998}, multi-level facility location \citep{Medrano-Gomez2020}, continuous coverage relationships \citep{Karasakal2004}, and modeling the upstream and downstream of industrial chains \citep{Zhu2010}. 

Moreover, with the availibility of large-scale trajectory data \citep{Liu2020}, these models have addressed the issue of static distance measurement in competition-based location choice. By using dynamic trajectories as the objects of coverage and constructing coverage relationships through stay points, departure points, or destinations, this model has expanded its applications in areas such as electric vehicle charging station locations \citep{Cai2014,Dong2014,Vazifeh2019} and remote workspace locations \citep{Caros2023}.

Nevertheless, trajectory-based coverage location choice studies generally treat trajectories as \emph{predefined inputs}. In reality, trajectories exhibit considerable randomness, and the papers using trajectories for location choices did not answer the question why trajectories offer a good foundation for location choices. In this paper we introduce a new perspective to theorize this contradiction: trajectories should be understood as \emph{samples} of human mobility, and thus, trajectory-based location choice can be viewed as targeting human mobility in the study area. If trajectories are well sampled, the location choice calculated should be considered representative. \footnote{For reference, given the many well-established laws of human mobility \citep{Gonzalez2008,Schlapfer2021,Song2010}, location choice based on good human mobility samples can achieve certain robust and even deterministic properties in a statistic sense. The variations in location choice outcomes due to changes in trajectory inputs are essentially a result of the randomness inherent in trajectory sampling.} We will build on this perspective in our subsequent research.

\section{Massive Retail Location Choice as a Set--Covering Problem}
\begin{figure}[!ht]
  \centering
 \includegraphics[width=3.1in]{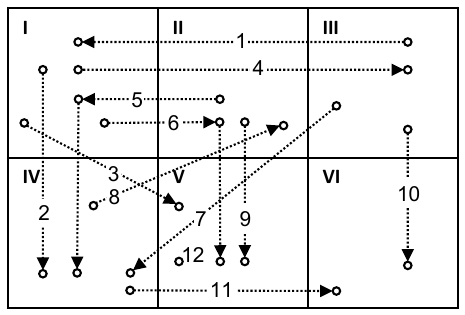}
  \caption{Diagram illustrating the trajectory--covering formulation.}
  \label{fig:diag}
\end{figure}

We will exemplify the notion of maximum customer covering through a simplified illustrative instance, clarifying the intuition as mentioned earlier. Figure \ref{fig:diag} presents this illustrative scenario, featuring six potential retail store locations and twelve customers. The customers' trajectories are depicted with dotted lines, and their ``stays'' in space are indicated by circles\footnote{In this context, the term ``stay'' is employed without explicit definition, relying instead on commonly understood interpretations. Broadly, it conveys a substantial period spent at a particular location, distinguishing it from mere transitory presence. Hence, in the illustrative case, only specific points on trajectories are designated by circles, while we consider all other points on the trajectories as instances of mere passing-by.}. The locations are denoted by Roman numerals, and customer trajectories are represented by Arabic numerals.

Supposing that every store location constitutes a \emph{suitably self-contained spatial unit}, exemplified by squares or designated wings within a mall, it ensures that any customer visiting the location will unavoidably be ``exposed'' to the retail store if it is established there. Under this premise, in the illustrative scenario, a store situated at location I will be capable of attracting the following ``exposed'' customers: 1, 2, 3, 4, 5, and 6, whereas location V will only capture customers 3, 9, and 12.

The primary aim of our massive location choice model is to maximize customer coverage within a fixed number of stores. Initially, the intuition suggests that this can be achieved by prioritizing locations with the highest number of ``stays'' and then proceeding to the next highest ones. At first glance, this approach appears promising in achieving our objective. However, the trajectories of customers can significantly influence the effectiveness of this seemingly straightforward ``na\"{\i}ve greedy'' solution, potentially affecting the overall solution quality.

Consider an extreme scenario where the majority of customers gravitate towards only two locations, such as the two most eminent landmarks in a city. Consequently, these two locations would be deemed as the most frequently ``stayed'' ones. Nonetheless, in the event of opening only two stores, placing them at these two locations would render the second store on each customer's trajectory entirely redundant, given our model's objective, as all customers have already been exposed once. An improved approach would involve selecting one of the two locations and adding an additional location, wherein new customers visit, thus enhancing the overall effectiveness of our solution.

Therefore, in order to maximize customer capture, it becomes necessary to consider the ``dynamic nature'' of their trajectories. Although this intention may initially appear convoluted, we will show that it can be simplified to a basic set-covering problem. We formally define the general case as follows.

Let us begin by defining the set of locations $\mathcal{S}$, where each $s\in \mathcal{S}$ represents a \emph{potential} store location. The trajectory of each customer $i$ consists of a sequence of locations where they ``stayed.'' Formally, the trajectory $T_i$ of customer $i$ is represented as a sequence of length $k_i$, with each entry in the sequence belonging to $\mathcal{S}$, or formally stated as:
$$T_i: \langle s_1, s_2, \ldots, s_{k_i} \rangle.$$
Given that our primary focus lies in whether customer $i$ is captured or not during any of their stays---disregarding the exact number of captures (as a single exposure suffices for soliciting a purchase in our assumption)---we can condense the sequence's information into a set of ``stays,'' 
\begin{equation}
    \sigma_i = \bigcup_{j=1}^{k_i}{\{s_j\}},
    \label{eq:sigma}
\end{equation}
in which we denote $\sigma_i$ as the set of stayed locations of customer $i$.

Conversely, based on this customer--location relationship, we can represent the set of \emph{covered customers for location $j$,} expressed formally as:
\begin{equation}
\pi_j = \left\{i|j\in \sigma_i\right\}.
\label{eq:pi}
\end{equation}

Thus, we have laid the foundation for our maximum customer covering problem, which bears fundamental \emph{isomorphism} to a classical maximum set cover problem, specifically, the max-$k$ cover problem \citep{Feige1998}. In this context, the locations serve as covering ``sets,'' while the customers represent the ``elements'' to be covered. The formal description of the customer--store version of the maximum set cover problem can be stated as follows: we aim to find a subset $S \subset \mathcal{S}$, subject to $|S| \leqslant k$, that maximizes the cardinality $\left|\bigcup_{\pi_j\in S} \pi_j\right|$.

In line with the conventional formulation of the maximum set cover problem, we can express the maximum customer covering problem also as an Integer Linear Programming (ILP) problem. The selection of each location is represented by a binary variable, denoted as $x_j = \mathbb{1}_{\sigma_j \in S}$. Additionally, the covering status of each customer is indicated by another binary variable, $y_i \in \{0, 1\}$. The relationship expressed in Eqs.\,(\ref{eq:sigma}--\ref{eq:pi}) allows us to define $y_i = \mathbb{1}_{\sigma_i\cap S \neq \varnothing}$, meaning that a customer is considered covered if any of the chosen locations intersect with the locations they stayed at.

The above relationship can be further simplified to the following optimization problem:

\begin{eqnarray}
\max &\sum\limits_i y_i, \nonumber\\
\mbox{s.t.}  &\sum\limits_j x_j \leqslant k, \nonumber\\
&  \sum\limits_{i \in \pi_j} x_i \geqslant y_j,   \nonumber\\
&x_i, y_j \in \{0, 1\}.\nonumber
\end{eqnarray}
The second condition implies $y_i = \mathbb{1}_{\sigma_i\cap S \neq \varnothing}$, signifying that a customer is not considered ``covered'' ($y_i = 0$) if none of the locations they stayed at is chosen as a store location.
\section{Approximation of Solution}
The maximum set cover problem is famously recognized as NP-hard, necessitating the adoption of approximation methods for solving the retail store location choice problem. In our specific case, as will be introduced shortly, we encounter tens of thousands of $x_j$'s and millions of $y_i$'s. To address this real-world scenario, our solution incorporates the following two approximation methods.
\begin{enumerate}
\item {\bf{Greedy:}} While we have scrutinized the ``na\"{\i}ve greedy'' approach for tackling the maximum coverage problem with customer trajectories, it is worth noting that in the well-established solution to the maximum coverage problem, a proven and efficient greedy method exists, capable of providing remarkably satisfactory solutions of high quality \citep{Hochbaum1996}. However, instead of na\"{\i}vely selecting the most ``stayed'' locations, i.e., those with the largest $|\pi_j|$, in a sequential manner, the more sophisticated greedy algorithm employs a different strategy. It chooses the most ``stayed'' location after deducting the customers already covered. Specifically, after the first $n$ locations $s_1, \ldots, s_n$ have been chosen, in the subsequent step, the approach selects the location $t$ that covers the most ``uncovered'' customers, formally expressed as:
$$s_{n+1} = \argmax_{s_j} |\pi_j \backslash \bigcup_{l=1}^n \pi_l |.$$
\item {\bf{Continuous approximation:}} Another conventional approximation approach worth mentioning involves treating $x_j$ and $y_i$ as continuous variables bounded between 0 and 1. After solving, the non-zero $x_j$'s are ranked, and the top $k$ are selected. It is important to note that while we will not further analyze the detailed mathematical properties of this approach, it is efficient, as it transforms into a straightforward linear programming problem, and yields satisfying solutions.
\end{enumerate}
\section{Data and Experiment}
We use Shenzhen, China, as a case to demonstrate the application of the aforementioned method, hypothetically for a real-world consumer electronics corporation which we selected. Providing context, Shenzhen had a resident population exceeding 17 million in 2022, and this corporation operated 129 stores in the city as of 2019.

We utilized a combination of the existing retail locations of the corporation and human mobility data for our analysis (refer to Table \ref{tab:data}). Specifically, the human mobility data was sourced from Baidu Huiyan\footnote{See \url{https://huiyan.baidu.com}.}. This dataset consists of entirely anonymized individual mobility trajectories, featuring a spatial resolution of up to 110 m $\times$ 110 m and a temporal resolution measured in seconds. The dataset encompasses the complete administrative area of Shenzhen, spanning the entire day of December 4, 2019.

\begin{table}
\footnotesize
  \centering
  \caption{Summary of data used in the paper}
  \begin{tabular}{p{2cm}p{5cm}p{1.8cm}p{2cm}p{2cm}}
    \toprule
    Dataset & Description & Spatial  Resolution & Data Collection Time & Source\\
    \midrule
    Population distribution&  Working, residential, and temporary visiting population in the analysis grids & 150--500 m
 & September, 2018 & Smartsteps \\
    Individual trajectory & Trajectories recorded when using Baidu Location Service & 150--500 m & December 4, 2019 & Baidu Huiyan\\

     Current retail location & Geocoded retail locations of the corporation in Shenzhen & 10 m & April, 2020 & Baidu Maps \\
    \bottomrule
  \end{tabular}
  \label{tab:data}
\end{table}

\begin{figure}[!th]
  \centering
  \subfigure[Locations of stays identified from the sample individual trajectories]{
    \includegraphics[width=0.8\textwidth, trim={0.5cm 83cm 20cm 0},clip]{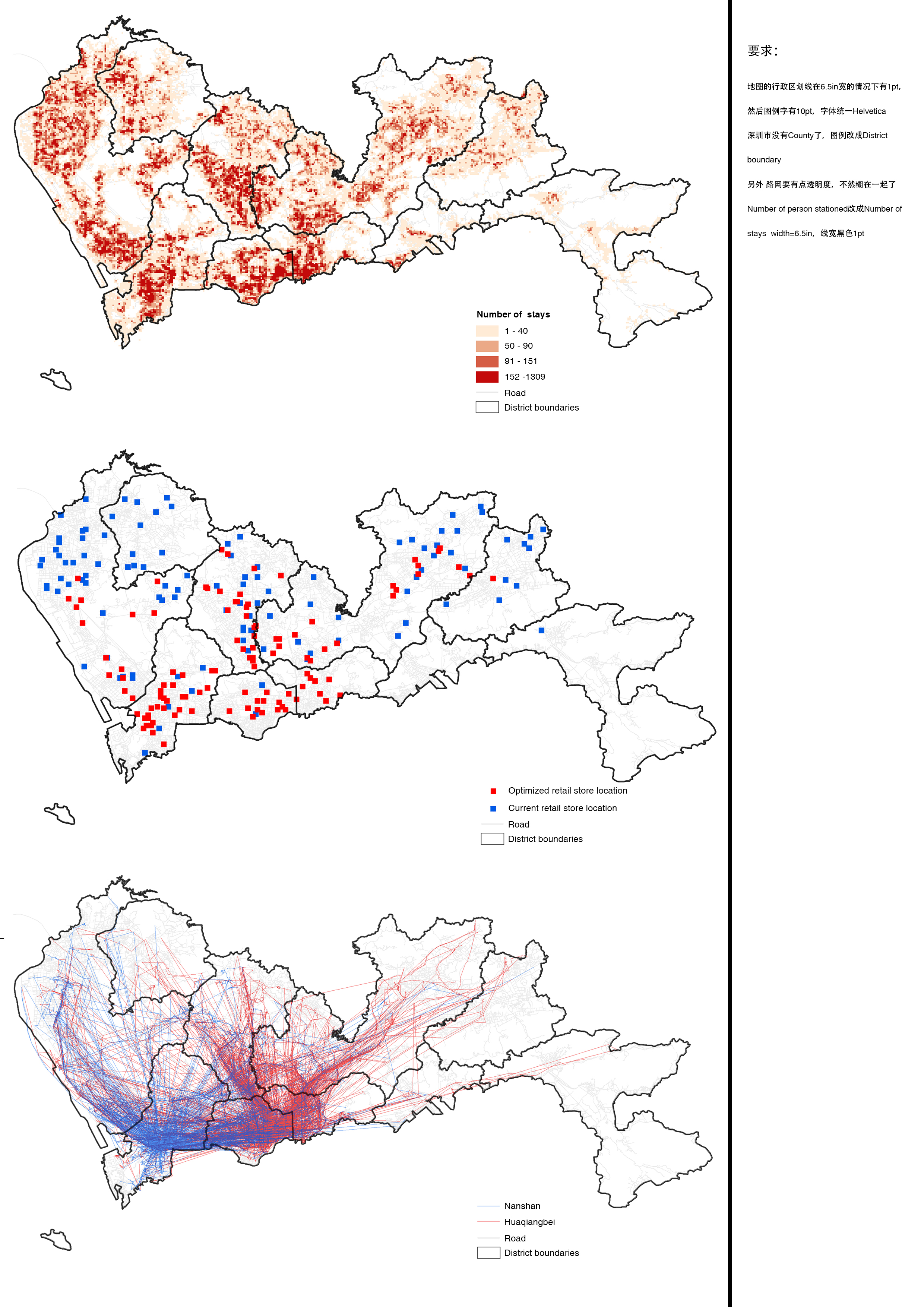}
    \label{fig:stay}
  }
  \subfigure[The trajectories of people who visitied \emph{N\'ansh\=an} and \emph{Hu\'{a}qi\'{a}ngb\v{e}i}]{
 \includegraphics[width=0.8\textwidth, trim={0.5cm 4cm 20cm 83cm 0},clip]{f1.jpg}
    \label{fig:twosites}
  }
  \caption{Trajectory data illustration for analysis in the model}
  \label{fig:data}
\end{figure}

The locations of stay identified using this data are shown in Figure \ref{fig:stay}. An illustration of the trajectories which passed two main business districts in Shenzhen, \emph{N\'{a}nsh\={a}n} and \emph{Hu\'{a}qi\'{a}ngb\v{e}i}, is shown in Figure \ref{fig:twosites}.

Furthermore, for acquiring fundamental demographic information, we utilized cellphone signaling data from China Unicom Smartsteps\footnote{See \url{http://www.smartsteps.com}.}. We used this dataset to estimate the working, residential, and transient visiting populations within the same grids.

\section{Results}
\begin{figure}[!ht]
  \centering
 \includegraphics[width=0.8\textwidth, trim={0.5cm 43cm  19cmcm 40cm},clip]{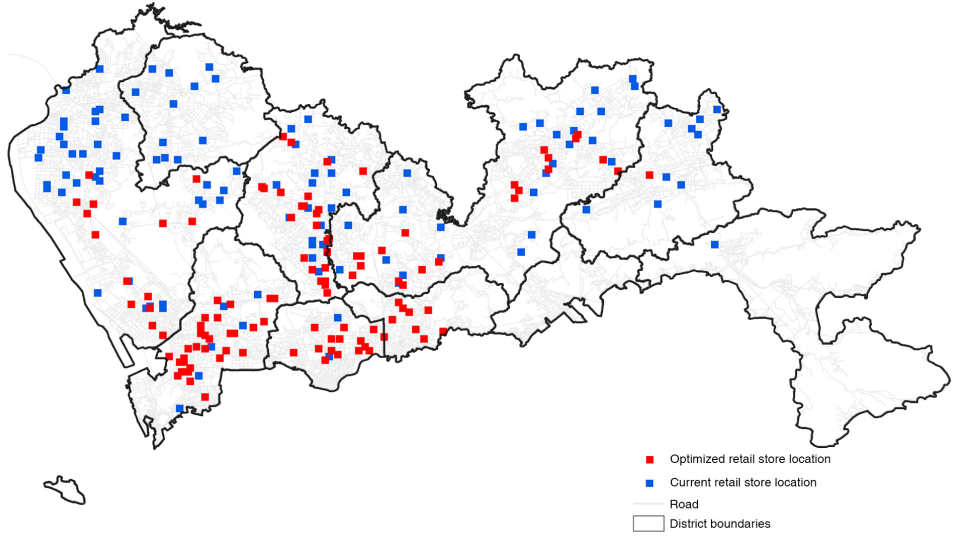}
  \caption{Location choice results in Shenzhen for the corporation}
  \label{fig:result}
\end{figure}

We select the target number of stores, setting $k=129$, to align with the existing store count. In Figure \ref{fig:result}, the outcomes of the location selection using the maximum coverage approach are presented, along with the current store locations for reference.

\begin{figure}[!ht]
  \centering
  \subfigure[Duplicate coverage included]{
    \includegraphics[width=0.45\textwidth, trim={20cm 87cm 20cm 3cm},clip]{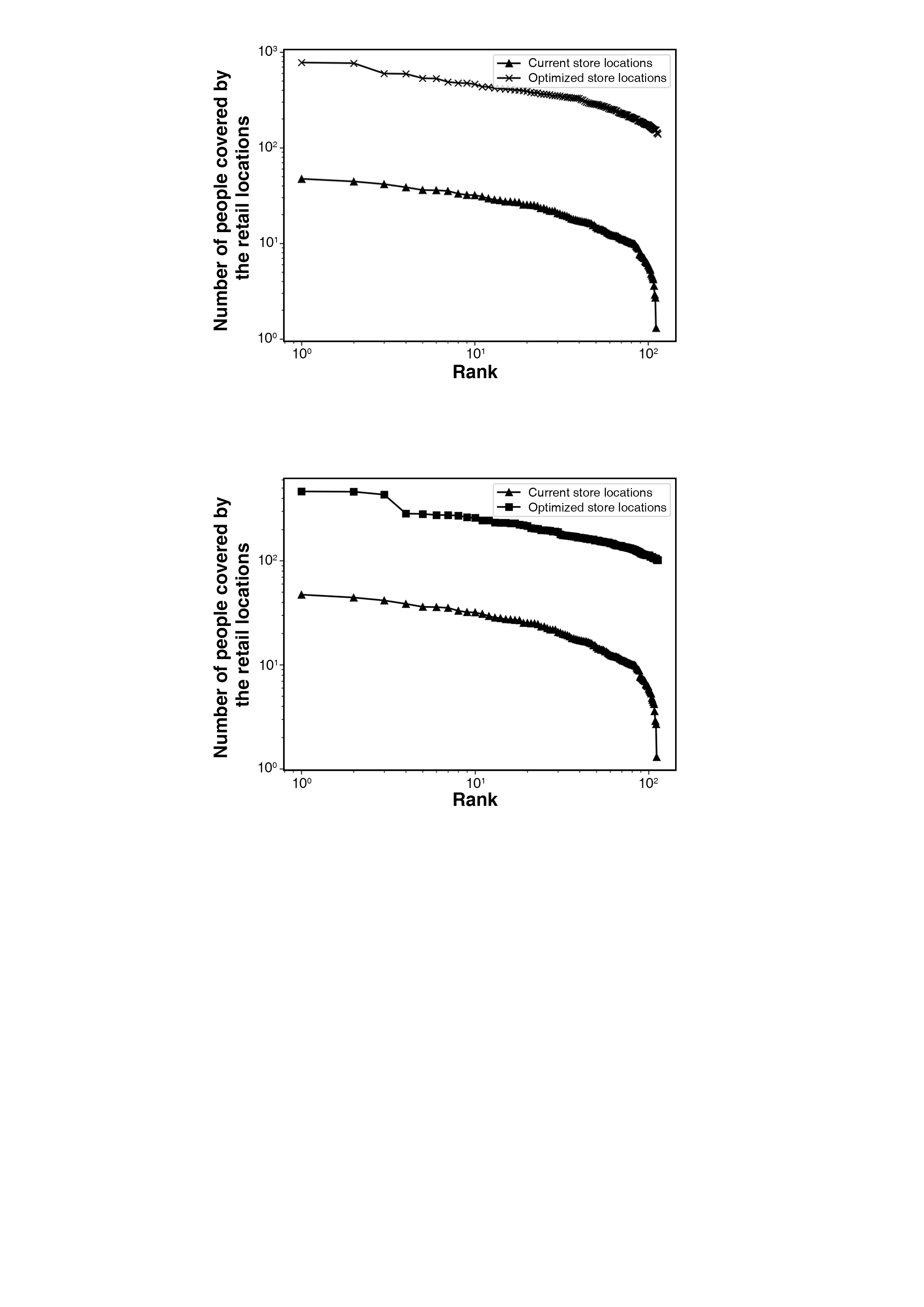}
    \label{fig:duplicate}}
  \subfigure[Duplicate coverage excluded]{
    \includegraphics[width=0.45\textwidth, trim={20cm 46.5cm 20cm 40cm},clip]{f2.jpg}
    \label{fig:noduplicate}}
  \caption{Evaluation of location choice results, with trajectory coverage for the current retail locations (triangles) and the optimized locations (asteroids/squares)}
  \label{fig:res_eval}
\end{figure}

In Figure \ref{fig:result} we may find that many of the optimized retail locations are actually closely situated to the existing ones, with some even overlapping. Nevertheless, when these optimized locations are ranked based on the population covered, as opposed to the existing stores, and plotted on log--log coordinates (see Figure \ref{fig:res_eval}), a distinct pattern emerges. The coverage of people  with the \emph{status quo} retail locations follows a curve with a declining tail, whereas the optimized locations exhibit a straight--line trend with increasing rank order.

Clearly, the two curves adhere to the power-law distribution (depicted as a straight line) and the truncated power-law distribution (exhibiting a declining tail), respectively. According to \cite{deluca2013fitting}, the corresponding equations are:
\begin{eqnarray}
N(r) &=& Cr^{-D}, \nonumber\\
M(r) &=& C(r^{-D}-r_T^{-D}), \nonumber
\end{eqnarray}
where $N$ and $M$ denote the proportion of the total population covered, while $r$ signifies the rank. In both equations, $D$ symbolizes the slope of the fitting line, commonly referred to as the power coefficient, and is fitted to be 0.31 in the Shenzhen case.

To explain the distinction between the two distributions, truncated power-law distributions typically arise from ``underdeveloped power-law distributions in the tails.'' In the context of the location choice problem, this phenomenon is likely attributed to poorly located stores with relatively small numbers of people to cover, creating a ever-enlarging gap between the achieved coverage and the optimal potential. In contrast, the power-law distribution of optimal results establishes a theoretical limit. This is further substantiated by the coverage rates: the trajectory coverage of existing store locations is 1.12\%, whereas the optimized coverage is 11.87\%, marking an almost ten-fold increase.

Moreover, the consistently observed power-law distribution in the rank--size distribution of stores' coverage suggests a plausible approach for determining a key hyperparameter in this study—the number of stores, denoted as $k$. This empirical regularity indicates diminishing efficiency gains with additional store openings concerning customer coverage. Consequently, instead of establishing a threshold based on the percentage of customer coverage, a more pragmatic approach in retail practice could involve setting a threshold for the volume/cost efficiency of new store openings in terms of customer coverage. This methodology offers an operational means of determining the optimal total store count for a retail chain within a single market area.

\section{Conclusion and Discussion}
In this paper, we introduced an optimization approach to address the mass location choice problem for a chain of retail stores. The objective is to maximize population coverage within a predefined number of stores. Our proposed methodology integrates individual human movement trajectories and employs a deduplication greedy algorithm. We illustrated the application of our method using Shenzhen as a case study, and the outcomes of the location choices demonstrate a significant improvement over the existing retail locations in terms of customer coverage. Furthermore, the optimized population coverage results exhibit a  power-law distribution. This finding not only enhances our understanding of the overall scaling effect in chain retail location potential, but also holds implications for future explanations in urban science.

Furthermore, our approach demonstrates practical utility in two distinct scenarios within the real-world context of retail location selection. First, in the context of establishing stores in a completely ``new town'' or a newly developed market area without pre-existing stores, our model stands suitable for comprehensive location selection covering the entire city or region. Second, the algorithm is also applicable for ``incremental optimization'' when augmenting a city's market with $k$ new stores that supplements existing ones. In this case, the identification and exclusion of trajectories covered by current stores, along with removing the already covered locations from potential solutions, form a crucial step. Leveraging the remaining trajectories, the algorithm achieves ``incremental optimization'' through a process termed as ``superimposed site selection.''

In alignment with the research gap outlined earlier, our proposed method effectively addresses challenges related to multicentricity of market areas, heterogeneous transportation costs, and computational complexities.

Future research can significantly expand the scope of this article in several dimensions: 1) The omission of online shopping's impact is a notable gap. Current evidence underscores the reliance of consumer preferences on firsthand product experiences, emphasizing the importance of physical retail stores, at least as display showrooms. A more comprehensive exploration of the interplay between online and offline shopping and its influence on store location choices warrants investigation. 2) The store location problem under customer segmentation presents another avenue for exploration. The current model, lacking trajectory attribute information, assumes the entire urban population as the target customer. In practical retail operations, specific customer groups with distinct characteristics are often considered. While privacy concerns constrained this study from incorporating individual attribute data, addressing these concerns would allow the straightforward introduction of such data. Given the spatial clustering tendency of similar population groups, targeting specific segments could enhance coverage outcomes. 3) An additional aspect is the determination of ``effective stay.'' This study defines one ``stay'' as effective individual coverage. However, in reality, multiple stays may increase the likelihood of purchase. Therefore, incorporating the number of stays for individuals into the model also represents a potential future direction. 4) While the current focus of this paper is on chain retail store location selection, the methodological framework is evidently transferable to other location tasks with population coverage objectives. For example, in outdoor \emph{advertising} location selection, where the model must consider coverage of pedestrian traffic, the basic structure mirrors that of retail store location optimization. The constraints and decision variables align, with the primary difference lying in the objective function. Store location selection aims to avoid repeated coverage of the target customer group, whereas outdoor advertising location selection seeks to enhance repeated coverage, reinforcing people's impressions of the advertisements. This approach is also  relevant for location problems pertaining to \emph{public infrastructure} with a focus on population coverage, such as the placement of public libraries or hospitals.

\section*{Acknowledgement}
This research is partially supported by the PKU--Wuhan Institute for Artificial Intelligence, PKU Institute of Urban Governance, and PKU Institute of Public Governance.
\bibliographystyle{apalike}
\bibliography{maxlocation}
\end{document}